\documentstyle[12pt]{article}
\begin{document}

\setcounter{page}{0}

\begin{flushright}
CERN-TH. 7429/94
\end{flushright}
\vspace{2.0cm}

\begin{center}
{\LARGE {\bf HIGH DENSITY QCD}}
\medskip

{\LARGE {\bf AND ENTROPY PRODUCTION}}
\medskip

{\LARGE {\bf  AT HEAVY ION COLLIDERS
\footnote{Talk presented at the NATO Advanced Research Workshop on
{\it Hot Hadronic Matter: Theory and Experiment},
Divonne, France,  July 1994.}
}}
\end{center}
\bigskip

\begin{center}

{\Large
{\bf K. Geiger}
}

{\it CERN TH-Division, CH-1211 Geneva 23, Switzerland}
\end{center}
\vspace{2.0cm}

\begin{center}
{\large {\bf Abstract}}
\end{center}
\medskip

The role of entropy production in the context of probing QCD properties at high
densities and finite temperatures in ultra-relativistic collisions of
heavy nuclei is inspected.
It is argued that the entropy generated in these reactions provides
a powerful tool to investigate the space-time evolution and the
question whether and how a deconfined plasma of quarks and gluons
is formed. I will address the questions how entropy is produced, and
how it is measurable. The uncertainties in predicting
the different contributions to the total entropy and particle multiplicities
during the course of heavy ion collisions are also discussed.

\noindent

\vspace{0.5cm}

\rightline{e-mail: klaus@surya11.cern.ch}
\leftline{CERN-TH. 7429/94,  September 1994}

\newpage

\noindent {\bf 1. INTRODUCTION}
\smallskip

In this talk I would like to present a  perspective
of what we can learn about aspects of QCD at high density with the advent
of a new generation of accelerators.
Particularly the future
heavy ion ($AA$) colliders, the BNL Relativistic Heavy Ion Collider (RHIC)
with maximum available beam energy $\sqrt{s}=200 \;A$ GeV (gold on gold) and
the CERN Large Hadron Collider (LHC) with $\sqrt{s}= 5500\;A$ GeV (lead on
lead),
will provide for the first time the opportunity
to study nuclear matter under extreme density compression
and very high temperatures, and possibly the formation of a deconfined
quark-gluon plasma (QGP). A major goal of the experimental programs at RHIC and
LHC
is to explore new phenomena associated with the dynamics of quarks and gluons
in the hot,
ultra-dense environment that is created in these collisions,
including the expected (non-) equilibrium QCD phase transition.
\bigskip

\noindent {\bf 2. THE REGIME OF HIGH DENSITY QCD}
\medskip

Fig. 1 shows the map of QCD as it is understood today. It exhibits three
very distinct regions of QCD with quite different physics \cite{levin94}.
On the horizontal axis,
$r$ characterizes the space-time distance that can
be resolved by probing QCD properties in certain dynamical processes.
For instance, when probing a nucleon or nucleus in a deep inelastic scattering
event with a momentum transfer $Q^2$, the photon acts as a microscope with a
resolution
$r \sim 1/Q$.  On the vertical axis,
the density $\rho_{qg}$ is the number of quark and gluon quanta
with a definite value of rapidity $y = \ln(1/x)$ that is
seen by our probe in the transverse plane,
\begin{equation}
\rho_{qg}\;=\;\frac{1}{\pi R^2}\,\frac{d N_{qg}}{dy} \;\simeq\;
\frac{A\, xf(x,Q^2)}{\pi R^2}
\;,
\end{equation}
where $f(x,Q^2)$ denotes the sum of quark and gluon structure functions,
$R$ is the nucleon radius and $A$ is the number of nucleons.

The three regions of Fig. 1 are:
\begin{itemize}
\item[(i)]
$r \ll 1$ fm and $\rho \ll R^{-2}$:
This is the small distance, low density
regime, where the powerful methods of perturbative QCD (pQCD) apply.
\item[(ii)]
$r \approx 1$ fm:
This is the non-perturbative QCD (npQCD) domain, where one has to deal
with the complex mechanisms of confinement and relys on lattice
calculations and QCD sum rules.
\item[(iii)]
$r \ll 1$ fm and $\rho \, \lower3pt\hbox{$\buildrel >\over\sim$}\, R^{-2}$:
Here we can probe a high density of partons at short distances (hdQCD).
Perturbative techniques can be applied, but in addition, one must
tackle important density effects that lead to interference and
collective phenomena.
\end{itemize}
\smallskip

The hdQCD regime is the exciting new field that opens up with the
experimental programs at HERA, RHIC and LHC.
There are two ways to obtain a system of partons with large density:
One way to penetrate  hdQCD is deep inelastic $ep$ scattering
($A=1$) at high energy in the region of very small Bjorken $x\ll 1$.
For instance, at HERA, the new experiments measure 30-50(!) gluons
in a proton at $x = 10^{-4}$ \cite{hera}.
The other access to the hdQCD region is through collisions of heavy nuclei,
in which one can reach high parton densities at not so very high energies
or small $x$, due to the large number of overlapping nucleons ($A\gg 1$).
This presumably can be achieved at RHIC ($x\approx 10^{-1}$), but certainly
at LHC ($x\approx 10^{-3}$). In particular at LHC both conditions, small $x$
and large $A$,
are combined.
\bigskip

\noindent {\bf 3. THE ROLE OF ENTROPY IN HEAVY ION COLLISIONS}
\medskip

For the remainder of this talk, I will focus on the hdQCD physics
in heavy ion collisions at RHIC and LHC.
Qualitatively, the  expected space-time evolution  of these reactions
can be summarized as illustrated in Fig. 2 \cite{msrep}:
(1) Immediately after the first nuclear contact,
the partons of the colliding nuclei begin to interact frequently with each
other,
resulting in a vehement materialization of excited quanta. (2) The excited
partons can rescatter and emit new particles and thereby evolve through
a  pre-equilibrium stage towards a thermalized quark gluon plasma
state, from which the system evolves further according to
the laws of relativistic hydrodynamics.
(3) The plasma expands and cools, and eventually a phase transition
- perhaps via a mixed parton-hadron phase -
into a purely hadronic gas occurs. (4) The freeze-out of this excited hadron
matter
produces the  final hadron yield.

One of the most valuable quantities for extracting information about
the dynamical evolution is the production of entropy. With each
of the above stages (1)-(4) a certain amount of generated entropy is
associated.
The entropy is generally defined in terms of the density matrix $\hat \rho$
of the quantum system, $S=-\mbox{Tr}\hat \rho \ln \hat \rho$ \cite{elze94},
or equivalently, in terms of the particle distribution (Wigner) function
$W(r,p)$, where $r\equiv r^\mu=(t,\vec r)$, $p\equiv p^\mu=(E,\vec p)$:
\begin{equation}
S(t)\;=\;-\,\int \frac{d^3r d^3 p}{(2\pi)^3} \,
W(r,p) \,\ln W(r,p)
\,,
\end{equation}
(neglecting Fermi or Bose statistics).
Thus $S$ contains valuable information about the time evolution of the system
in full phase-space,
with the complete history up to time $t$ embodied in the function $W$.
Measuring the entropy as a function of time would therefore allow
to trace the space-time evolution of a nuclear collision. Furthermore
the rate of entropy production reflects the degree and time scale of
equilibration, since $dS/dt\rightarrow 0$ as an equilibrium state is reached
\cite{degroot}.

Clearly, it is not easy to extract the time evolution of $S$ from experiment.
First of all the entropy is not directly observable, it can be
measured only indirectly through the multiplicities of produced particles
in a reaction and the amount of transverse energy generated.
Secondly, to measure the rate of entropy or particle production,
one has to identify signals of characteristic particle emission from different
stages of a nuclear collision. Good probes of time rate of change are
dileptons, photons, strange and charm production \cite{qm93,msrep}.
To this end, let me pose the following two questions:
\smallskip

{\it How is entropy  produced in heavy ion collisions at RHIC and LHC}?
\smallskip

{\it How can we measure entropy in these reactions}?
\smallskip
\bigskip

\noindent {\bf 4. HOW IS ENTROPY PRODUCED?}
\medskip

Most of the entropy and transverse energy is expected to be
produced already within the first $fm$
after nuclear contact by very frequent so-called {\it semihard}
parton interactions with only a few GeV momentum transfer \cite{glr,msqm93}.
This corresponds to the pre-equilibrium regime in Fig. 2.
Once the system has reached a thermal equilibrium state, the entropy
production vanishes in space-time, and ideally the final transition from the
quark-gluon phase to the hadron phase and its freeze out should keep
the entropy constant \cite{bj83}. Let me write the total produced entropy
that is produced during the course of a heavy ion collision as
\begin{equation}
\Delta S \;=\; \Delta S^{(qg)}_{prim} \;+\; \Delta S^{(qg)}_{sec} \;+\; \Delta
S^{(had)}
\,.
\end{equation}
The three contributions arise as follows:
\begin{itemize}
\item[(i)]
$\Delta S_{prim}^{(qg)}$, {\it large}:
primary parton production  due to
the materialization of virtual "pre-existing" quanta in the colliding nuclei
that are set free by very frequent initial parton scatterings.
\item[(ii)]
$\Delta S_{sec}^{(qg)}$, {\it very large}:
secondary parton production due to
intense gluon bremsstrahlung off partons that have been excited in scatterings,
plus production by rescatterings which become increasingly probable as the
parton density grows.
\item[(ii)]
$\Delta S^{(had)}$, {\it small}:
entropy produced through the parton-hadron transition (which ideally
should be neglegible) plus additional entropy generated by decay of
formed "pre-hadrons" and hadronic resonances into the final stable
hadron states.
\end{itemize}

As an illustrative example,
Fig. 3a displays the time development
of the specific entropy $(S/N)(t)$ arising from $\Delta S_{prim}^{(qg)}$
and  $\Delta S_{sec}^{(qg)}$
as calculated within the parton cascade model \cite{msrep} for
various collider energies.
The curves show a rapid build-up of $S/N$ and relax approximately
exponential to reach their final values between 3.9 and 4.3.
Comparing these values with $(S/N)_{ideal} \simeq$ 4
for an ideal gas of non-interacting massless quarks and gluons, one
sees that the difference between the resulting entropy of the realistic
model calculation and the idealized case amounts
less than 10 \%.
Although the model includes massive quarks and accounts for interactions
among the partons,
the system of partons behaves effectively like an almost ideal gas.
Fig. 3b shows the corresponding relaxation times versus $\sqrt{s}/A$, which
are evidently very short and indicate a very rapid thermalization (plasma
formation) within less than 0.5 $fm$.
\bigskip

\noindent {\bf 5. HOW CAN ENTROPY BE MEASURED?}
\medskip

As mentioned above, the entropy ideally must stay constant once the system has
reached
an equilibrium (QGP) state. In this context, the only effect of the
parton-hadron phase transition is
a reorganization of the degrees of freedom from the colored quarks and gluons
to the color singlet hadrons, mostly pions.
Now, experiments of deep inelastic $ep$-scattering, $e^+e^-$-annihilation,
Drell Yan, etc.,
strongly support the hypothesis of local parton-hadron duality \cite{lphd}.
That is, it appears that in high energy processes the mechanism of
hadronization is a universal and local
phenomenon, independent of the partons prehistory. The measured hadron
multiplicity
turns out to be simply equal the calculated parton multiplicity times a
constant
(e.g. $N^{(\pi)} = 1.1 N^{(qg)}$).

On the other hand, if there is a first order QCD phase transition
at a temperature $T_c$,
the ratio of the entropy density
in the pion plasma, $s^{(\pi)}$, to that of the
quark-gluon plasma, $s^{(qg)}$, can be estimated from the effective number of
degrees of freedom in the two phases at $T_c$ as
$ r = s^{(\pi)}/s^{(qg)}\approx 0.7$ \cite{bj83}.
Under these assumptions for zero impact parameter collisions
the total produced entropy can be measured by
\begin{equation}
\frac{d S}{d y} \,=\,
c^{(qg)} \,\left( \frac{d N^{(qg)}}{d y}\right)_{b=0} \,
\,\approx\,
\frac{c^{(\pi)}}{r} \, \left( \frac{d N^{(\pi)}}{d y}\right)_{b=0}
\;.
\end{equation}
where $c^{(qg)}\simeq 4$ and $c^{(\pi)}=3.6$.
Thus, again one has
$N^{(\pi)} = c^{(qg)}/c^{(\pi)} N^{(qg)} = 1.1 N^{(qg)}$. Using the above value
for $r$,
one gets a relation between the  entropy produced per unit rapidity
and the multiplicity of final pions \cite{ms2}:
\begin{equation}
\left(\frac{d N_\pi}{d y}\right)_{b=0} \;\simeq\;
 0.2\; \frac{d S}{d y}
\;.
\end{equation}
The total produced entropy contained in an interval $d y$ around
$y=0$ is effectively related to the observable pion multiplicity
$d N^{(\pi)}/ dy$. Therefore the multiplicity of pions reflects
the entropy produced by the partons in the central collision region,
provided that the entropy production associated with
the freeze out of the excited hadron matter after the phase transition
and the decay chain through resonance states is marginal.
This is illustrated in Fig. 4.
\bigskip

\noindent {\bf 6. UNCERTAINTIES}
\medskip

There are essentially two sources of uncertainties in predicting
the amount of entropy and global observables as particle multiplicity or
transverse energy:

{\bf a)} The initial parton distributions in nucleus-nucleus collisions are
subject to
large uncertainty.
Particularly the small $x$ behaviour of the gluon structure functions and the
question of gluon shadowing \cite{nucshad2} in nuclei are of primary
importance, since
the amount of soft gluons in the initial state sensitively determines
the starting conditions for the evolution and of the general
reaction dynamics \cite{eskola94}.
These gluons densely populate phase-space and due to the large
global semihard cross-section
they only need to be slightly "tickled" in order to materialize.
Thus the gluon content
of the nuclei is a large entropy reservoir that is immediately set free because
the quantum coherence is so easily destroyed \cite{elze94}.
Even though the recent measurements at HERA \cite{hera} of the nucleon
structure functions
at very small values of $x$ with an extracted  gluon component
$x g(x)\propto 1/\sqrt{x}$ provide
new information, the question of the corresponding
form of the structure functions of nuclei composed
of many nucleons requires much further investigation.
Moreover, the measured structure functions, which contain solely the single
particle aspect of
the parton distributions, do not tell us anything about the spatial
distribution of partons, or
correlations among them - issues which presumably
play an important role for large nuclei.
Fig. 5 shows the impact of structure function dependence on the parton
production cross-section
in $Au+Au$ collisions at RHIC and LHC energies versus momentum cut-off $p_0$
\cite{eskola94}.
Most parton scatterings are semi-hard processes with momentum transfers $\simeq
2 - 3$ GeV, thus
the dominant region for particle production is at low $p_0$.
The variation with the choice of structure functions is especially drastic at
LHC,
where it reaches up to an order of magnitude difference in the cross-section.
\smallskip

{\bf b)}
The other  source of substantial uncertainty originates
from the current lack of better knowledge of the impact of
nuclear and dense medium effects which are absent
in e.g. hadronic collisions, but become important for heavy ion collisions.
The new physics associated with the nuclear and medium effects
on the parton level \cite{lbl93} that determine the dynamics of $AA$ collisions
drastically evident in  Fig. 6.
where the various curves a)-e) exhibit the characteristics of  different
evolution scenarios of simulations of gold on gold collisions \cite{ms3},
namely,
a) `naive evolution',
b) plus nuclear shadowing,
c) plus parton fusion and absorption,
d) plus Landau-Pomeranchuk-Migdal effect,
e) plus soft gluon interference.
Shown are  the charged particle distributions
in pseudorapidity ($\eta$),
calculated for central $Au + Au$ collisions with
$\sqrt{s} =$ 200 A GeV and $\sqrt{s} =$ 6300 A GeV.
Evidently, the particle density
in the central rapidity region is reduced successively from the `naive'
calculation (case a) to the default result (case e) by more than a
factor of 3.
\bigskip

\noindent{\bf 7. SUMMARY}
\medskip

Much  experimental and theoretical study is required
before robust conclusions for experimental observables and
open theoretical questions can be drawn -
although I believe, already at this point a
clear understanding
of the fundamental problems on the basis of perturbative QCD
and transport theory has been reached \cite{msrep}.
Foremost, we must learn how to
utilize medium effects in  high density QCD,
so that truly quantitative calculations become possible.
Almost as importantly, we must develop a microscopic theory for the
hadronization of a quark-gluon plasma that is based on QCD, doing away
with the presently employed QCD-inspired models that have little
predictive power in the new high density regime.
If these two goals can be met, quantitative
ab-initio calculations of ultrarelativistic heavy ion collisions will
become feasible by the time when RHIC and LHC start operation.

\newpage

{\bf FIGURE CAPTIONS}
\bigskip

\noindent {\bf Figure 1:}
The map of QCD with $\rho_{qg}$ denoting the densities of partons in
the transverse plane and $r$ the distance resolved in an experiment.
\bigskip

\noindent {\bf Figure 2:}
Space-time diagram of the (longitudinal) evolution of a
relativistic nucleus-nucleus collision.
\bigskip

\noindent {\bf Figure 3:}
Results of a Monte Carlo simulation of gold on gold collisions at
different beam energies.
a) Entropy per secondary particle produced as a function of real time.
b) Corresponding relaxation times versus beam energy.
\bigskip

\noindent {\bf Figure 4:}
Schematics of the entropy production in the parton and hadron phases
during the evolution of  nucleus-nucleus collisions.
\bigskip

\noindent {\bf Figure 5:}
Estimated parton production cross-section in heavy ion collisions at RHIC
and LHC for differents sets of structure function parametrizations
(from Ref. \cite{eskola94}).
\bigskip

\noindent {\bf Figure 6:}
Impact of nuclear and dense matter effects on a) the parton evolution
and b) the resulting charged particle spectra in
$Au+Au$ collisions at RHIC and LHC energies.
The different evolution scenarios are explained in the text.
\bigskip

\vfill
\end{document}